\pdfoutput=1
\documentclass{PoS}
\usepackage{amsmath,amssymb,bbm,color}
\usepackage[pdftex]{graphicx}
\usepackage{epsfig}
\usepackage{euscript}
\usepackage{pslatex}
\usepackage{fancybox}
\usepackage{enumerate}
\usepackage{widetext}
\usepackage{subfigure}
\def\rQCED{{\rm QCED}}
\newcommand{\KK}{${\cal KK}$}
\newcommand{\bhlumi}{{\tt BHLUMI}}
\title{Role of IR-Improvement in Precision LHC/FCC
Physics and in Quantum Gravity}

\ShortTitle{Role of IR-Improvement}

\author{\speaker{B.F.L. Ward}\\
        Baylor Un iversity, Waco, TX, USA\\
        E-mail: \email{bfl\_ward@baylor.edu}}

\author{S. Jadach\\
 Institute of Nuclear Physics, Krakow, Poland\\
   E-mail: \email{Stanislaw.Jadach@cern.ch}}

\author{W. Placzek\\
 Institute of Physics, Jagiellonian University, Krakow, Poland\\
   E-mail: \email{Wiesiek.Placzek@cern.ch}}

\author{M. Skrzypek\\
 Institute of Nuclear Physics, Krakow, Poland\\
   E-mail: \email{Maciej.Skrzypek@cern.ch}}

\author{Z. A. Was\\
 Institute of Nuclear Physics, Krakow, Poland\\
   E-mail: \email{Z.Was@cern.ch}}
    
\author{S.A. Yost\\
 The Citadel, Charleston, SC, USA\\
    E-mail: \email{yosts1@citadel.edu}}    

\abstract{IR-improvement based on amplitude-level resummation allows one to control unintegrable results in quantum field theory with arbitrary precision in principle. We illustrate such resummation in specific examples in precision LHC and FCC physics and in quantum gravity.}

\FullConference{14th International Symposium on Radiative Corrections (RADCOR2019)\\ 
9-13 September 2019\\
		Palais des Papes, Avignon, France}

\centerline{\bf BU-HEPP-19-08, Dec. 2019}
\begin{document}
\section{Introduction}
With the era of precision LHC physics squarely upon us, the relevance of resummation in quantum field theory is large. This relevance is amplified by the need to plan for future colliders in view of the current lack of experimental evidence for new degrees of freedom beyond those in the Standard Theory\footnote{We follow Prof. D.J. Gross~\cite{djg-smat50} and henceforth refer to the Standard Model as the Standard Theory of elementary particles.} in the LHC data at this writing. Indeed, several proposals for the next big colliding beam device involve a precision physics $e^+e^-$ colliding beams phase in the regime of $\sqrt{s}$ from $91$ GeV to 390 GeV as a first step: FCC, ILC, CEPC, and CLIC. We mention that, in the problem of the union of quantum mechanics with gravity as given by the Einstein-Hilbert theory, precision theory methods seem also to be relevant. In what follows, we discuss the role of IR-improvement in the particular approach amplitude-based resummation in quantum field theory as it is applied to the aforementioned precision physics considerations.\par
In discussing the role of IR-improvement in quantum field theory in precision physics considerations, we are called to address the general idea of resummation in quantum field theory for
IR, UV or CL (collinear) effects. The matter can be put into focus by comparing with the simple example of summation for a geometric series, such as
\begin{equation}
\frac{1}{1-x} = \sum_{n=0}^{\infty}x^n.
\label{eq-elmtry}
\end{equation} 
Here the result of the summation is an analytic result that is well defined except for a simple pole at $x=1$. While the mathematical tests for convergence on the simple series would only guarantee convergence for $|x| <1$, the result of the summation yields a result that is well-behaved in the entire complex plane except for a simple pole at $x=1$. The message is that infinite order summation can yield results that are much better behaved than what one can deduce order-by-order in the respective series. One is accordingly invited, as a next step, to resum series that are already being summed to try to improve ones knowledge of the represented function.\par
Continuing in this way, we write for the Feynman series for a process the following:{\small
\begin{equation}
\sum_{n=0}^{\infty}C_n \alpha_s^n \begin{cases}&= F_{\rm RES}(\alpha_s)\sum_{n=0}^{\infty} B_n \alpha_s^n,\; \text{\rm EXACT}\\
                                                                                                                                      &\cong  G_{\rm RES}(\alpha_s)\sum_{n=0}^{\infty} \tilde{B}_n \alpha_s^n,\; \text{\rm APPROX.}\end{cases}
                                                                                                                                      \label{eq-res}
\end{equation} }
Two versions of resumming the original Feynman series for the quantity on the LHS are given on the RHS. One is an exact re-arrangement of the original series. It is labeled `exact'.The other only agrees with the LHS to some fixed order N in the expansion parameter $\alpha_s$. It is labeled `approx'. Which version of resummation is to be preferred has been discussed for some time~\cite{frits-ichep88}. At this writing, the discussion continues to new paradigms.\par
In QCD$\otimes$ QED the discussion continues with added dimensions. We have the ongoing comparison, by two of us (BFLW and SAY), of the hard IR cut-offs in MC's such as Herwig~\cite{HERWIG}, Herwig++~\cite{Bahr:2008pv}, Pythia~\cite{Pythia} and Sherpa~\cite{sherpa} versus the resummed integrability of the IR in Herwiri1.031~\cite{hwiri}. In addition, we have the approaches to ISR radiation from quarks in hard hadron - hadron scattering processes, wherein we have the approach with QED PDF's~\cite{qedpdf} versus exact Feynman diagrams with short-distance quark masses as pursued by four of us (SJ, BFLW, ZAW, and SAY) in Ref.~\cite{kkmchh}. In the FCC and other future collider frameworks, the attendant precision physics programs necessitate IR-improved quantum field theory results to exploit properly the respective new physics windows.
In quantum gravity, it has been found that resummed IR leads to UV finiteness~\cite{reutera,laut,reuterb,reuter3,litim,litim1,litim2,perc,perc1,perc2,perc3,perc4,bw1rqg,bw2rqg,bw2arqg,bw2irqg}. In what follows, we expand on these developments in IR-improvement of quantum field theory from the standpoint of applications to precision LHC/FCC physics and applications of one of us (BFLW) of quantum gravity.\par
\begin{figure}[h]
\begin{center}
\setlength{\unitlength}{1in}
\begin{picture}(6,3.9)(0,0)
\put(0.5,0.2){\includegraphics[width=5in]{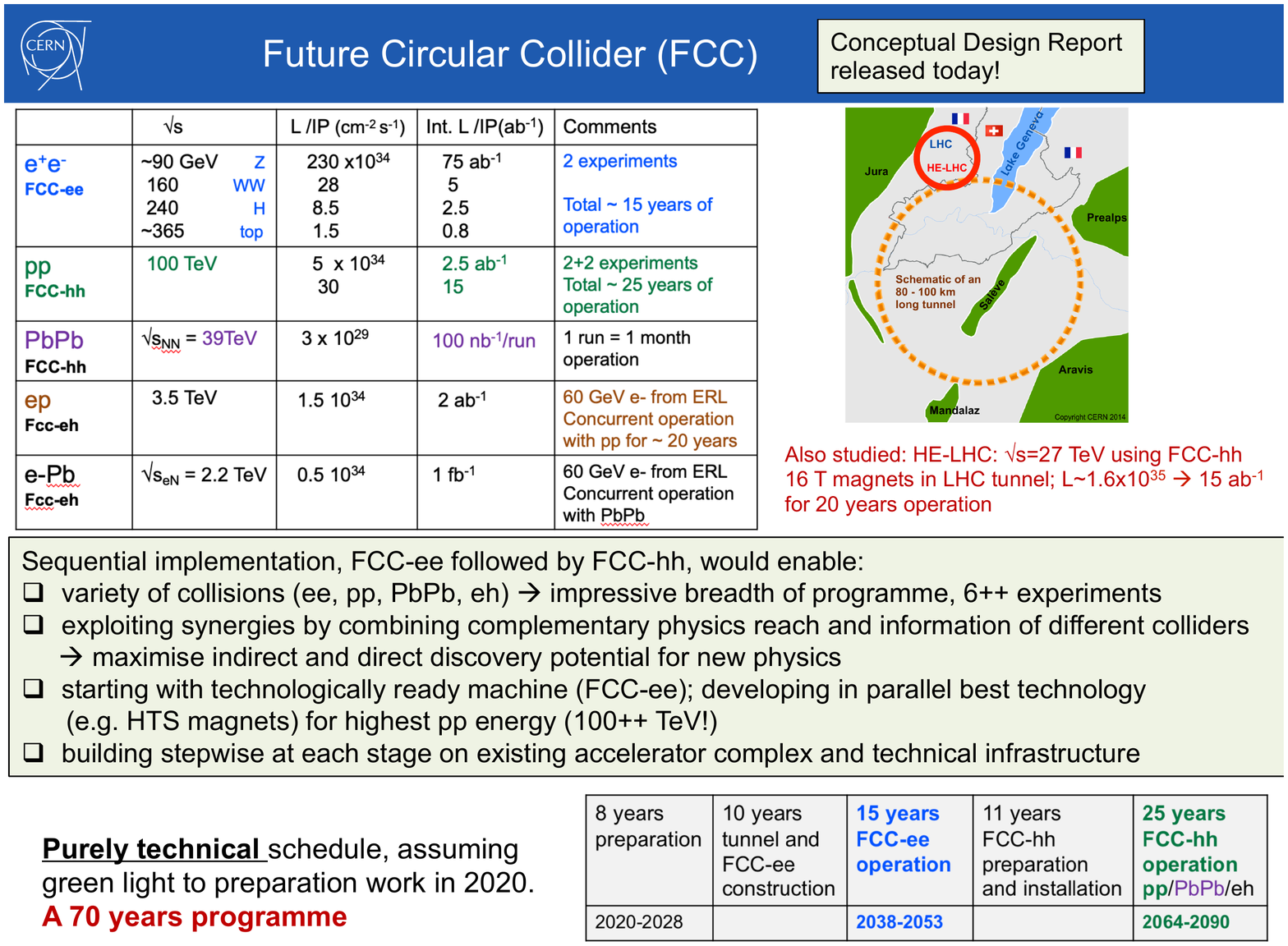}}
\end{picture}
\end{center}
\vspace{-10mm}
\caption{\baselineskip=11pt Tentative schedule and physical layout in FCC Study as portrayed in Ref.~\cite{fcc-fabiola-2019}.}
\label{fig1}
\end{figure}

We close this introductory section by calling attention to the 50-year celebrations of the Standard Model of Elementary Particles~\cite{SM3,SM4,SM1,BW1a,BW1b,BW1c,case-sm50,slacsi2018}. At the same time, plans are underway for the future explorations in our field as we have noted. We call attention here to the FCC Study~\cite{fcc-fabiola-2019} which we show in Fig.~\ref{fig1}.
The plans call for a precision studies  $e^+e^-$  phase featuring more than 5 Tera Z's, for example, followed by a 100 TeV hadron collider, reminiscent of the canceled SSC in Waxahachie, TX, USA. Only now do we really see how big of a mistake the latter cancellation was. It is important to keep this historical perspective in mind.\par
The discussion proceeds as follows. In the next section, we review briefly exact amplitude-based resummation theory. This is followed by a discussion of applications in precision LHC physics in Section 3. Applications in Precision FCC physics are discussed in Section 4. Section 5 present applications in quantum gravity. Section 6 contains our summary remarks.\par
\section{Review of  Exact Amplitude-Based Resummation Theory}
This approach to higher order corrections is founded on the master formula{\small
\begin{eqnarray}
&d\bar\sigma_{\rm res} = e^{\rm SUM_{IR}(QCED)}
   \sum_{{n,m}=0}^\infty\frac{1}{n!m!}\int\prod_{j_1=1}^n\frac{d^3k_{j_1}}{k_{j_1}} \cr
&\prod_{j_2=1}^m\frac{d^3{k'}_{j_2}}{{k'}_{j_2}}
\int\frac{d^4y}{(2\pi)^4}e^{iy\cdot(p_1+q_1-p_2-q_2-\sum k_{j_1}-\sum {k'}_{j_2})+
D_\rQCED} \cr
&{\tilde{\bar\beta}_{n,m}(k_1,\ldots,k_n;k'_1,\ldots,k'_m)}\frac{d^3p_2}{p_2^{\,0}}\frac{d^3q_2}{q_2^{\,0}},
\label{subp15b}
\end{eqnarray}}
where {\em new} (YFS-style) {\em non-Abelian} residuals 
{$\tilde{\bar\beta}_{n,m}(k_1,\ldots,k_n;k'_1,\ldots,k'_m)$} have {$n$} hard gluons and {$m$} hard photons. The infrared functions ${\rm SUM_{IR}(QCED)}$ and ${ D_\rQCED}$ and the residuals are defined in Ref.~\cite{mcnlo-hwiri,mcnlo-hwiri1}.  In the implementation of parton shower/ME matching,  we have the replacements {$\tilde{\bar\beta}_{n,m}\rightarrow \hat{\tilde{\bar\beta}}_{n,m}$}. These replacements  allow us, via the basic formula{\small
\begin{equation}
{d\sigma} =\sum_{i,j}\int dx_1dx_2{F_i(x_1)F_j(x_2)} d\hat\sigma_{\rm res}(x_1x_2s),
\label{bscfrla}
\end{equation}}
to connect with  MC@NLO~\cite{mcnlo,mcnlo1} as explained in Ref.~\cite{mcnlo-hwiri,mcnlo-hwiri1}.\par
We have made various applications of Eq.(\ref{subp15b}) to precision LHC and FCC physics and we have made applications of its extension to general relativity as an approach to quantum gravity. We discuss such applications in precision LHC physics in the next Section.\par 
\section{Applications in Precision LHC Physics}
In Refs.~\cite{mcnlo-hwiri,mcnlo-hwiri1,mg5_amc-hwri}, two of us (BFLW and SAY) have interfaced the realization of IR-improved DGLAP-CS ~\cite{dglap1,dglap2,dglap3,dglap4,dglap5,dglap6,dglap7,dglap8,dglap9} theory
as realized in Herwiri1.031~\cite{hwiri} with MC@NLO and with MG5\_aMC@NLO~\cite{mg5amcnlo} for precision  studies of Z and W+jets production at the LHC.
In Refs.~\cite{kkmchh,kkmchh1,radcor17} four of us (SJ, BFLW, ZAW, and SAY) 
have realized exact ${\cal O}(\alpha^2 L)$ CEEX EW corrections for hadron-hadron scattering processes in the MC {\KK}MC-hh 
which features built-in parton showers from Herwig6.5 and Herwiri1.031 as well as an LHE~\cite{lhe-formt} format to facilitate the interface with other parton shower MC's. 
In Z and W+ jets productions, IR-improvement gives a comparable or better data fit without ad hoc parameters for the intrinsic transverse momentum spectrum of partons inside the proton.
In {\KK}MC-hh, IR-improvement allows to quantify role of ISR in precision predictions for Z production observables, as we now illustrate.\par
\begin{figure}
\begin{center}
\setlength{\unitlength}{1in}
\begin{picture}(6,2.4)(0,0)
\put(0.0,0.2){\includegraphics[width=6in]{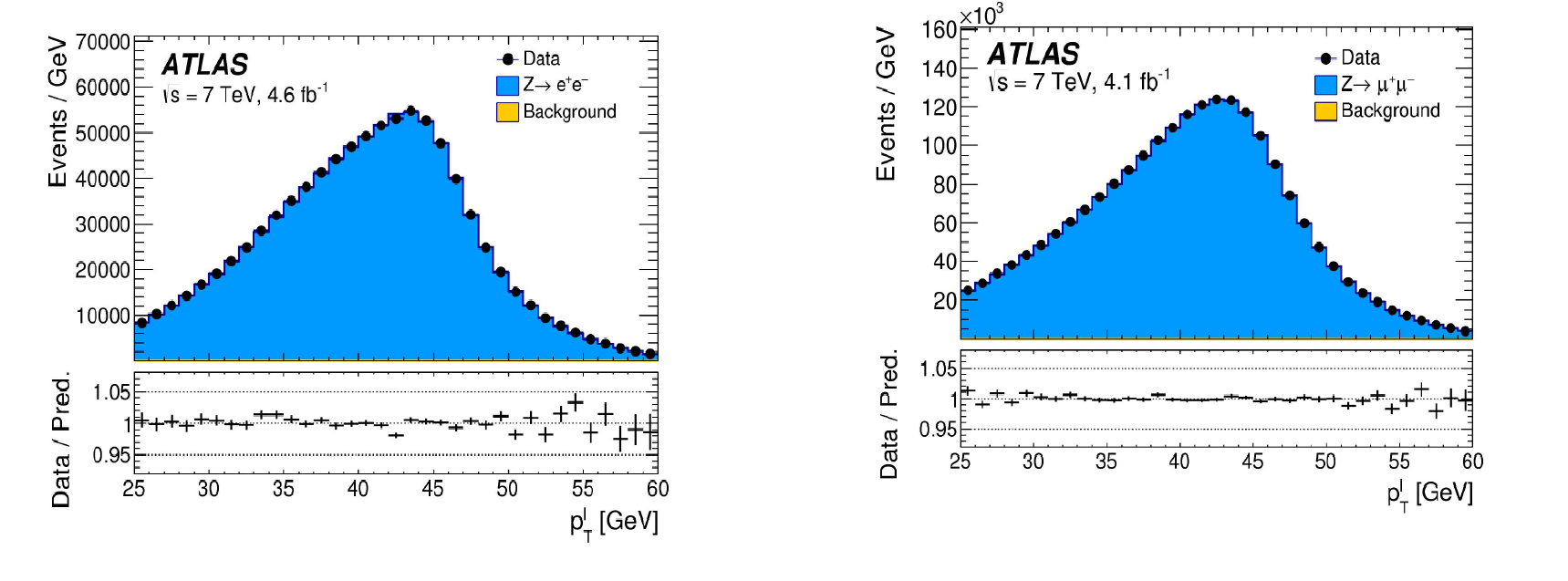}}
\end{picture}
\end{center}
\vspace{-10mm}
\caption{\baselineskip=11pt ATLAS electron and muon $p_T$ spectra from Fig. 15 in Ref.~\cite{atlasmw-17}.}
\label{fig2}
\end{figure}
\begin{figure}[h]
\begin{center}
\setlength{\unitlength}{1in}
\begin{picture}(6,2.4)(0,0)
\put(0,0.2){\includegraphics[width=3in]{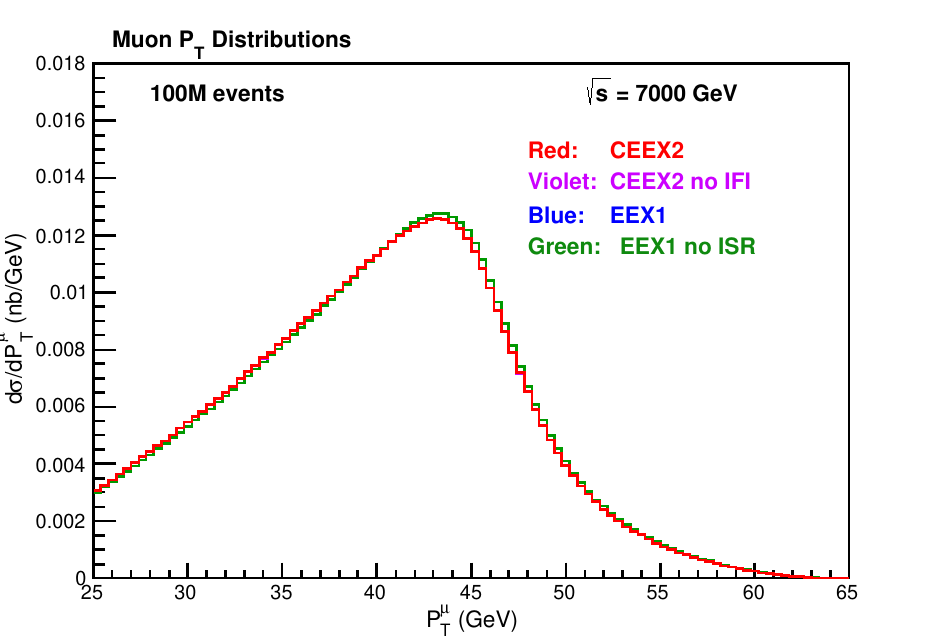}}
\put(3,0.2){\includegraphics[width=3in]{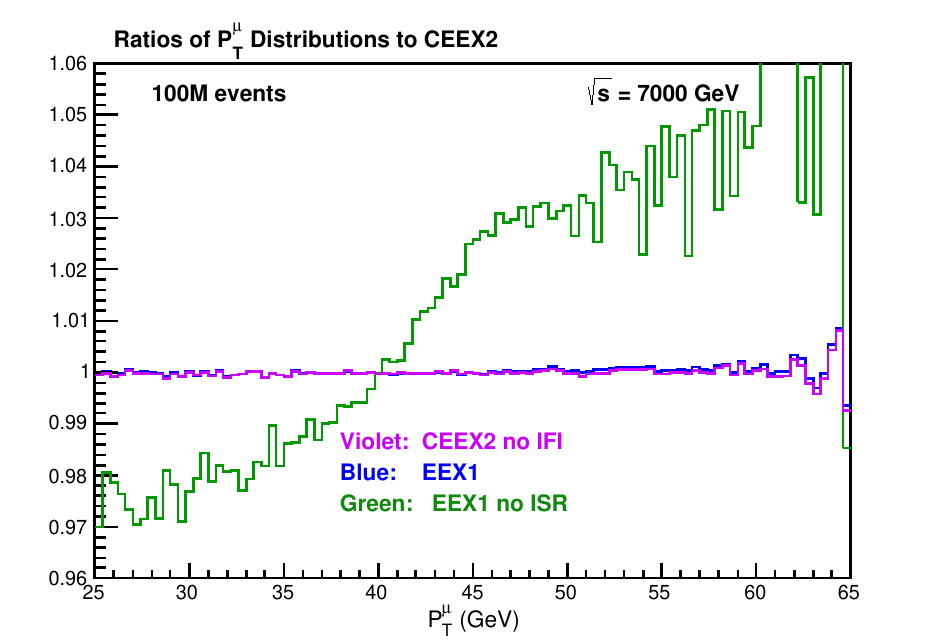}}
\end{picture}
\end{center}
\vspace{-10mm}
\caption{\baselineskip=11pt Muon transverse momentum distributions and their ratios for {\KK}MC-hh with the cuts specified in the text for the EW-CORR (electroweak-correction) labels ``CEEX2" (red -- medium dark shade), ``CEEX2 (no IFI)" (violet -- light dark shade), ``EEX1'' (blue -- dark shade), and ``EEX1 (no IFI)'' (green -- light shade), showered by HERWIG 6.5. The labels are explained in the text. The ratio plot features ``CEEX2'' as the reference distribution as noted in the respective title.}
\label{fig3}
\end{figure}

Specifically, our implementation for DIZET~\cite{zfitter1,zfitter6:1999} in {\KK}MC-hh uses a modified $G_\mu$ scheme with Standard Theory\footnote{We follow D.J. Gross~\cite{djg-smat50} and henceforth refer to the Standard Model~\cite{SM3,SM4,SM1,BW1a,BW1b,BW1c} as the Standard Theory.} input parameters as given in Ref.~\cite{vicini-wack:2016} with the exception that we use the light quark masses 
$m_u=6.0$ MeV,$\; m_d=10.0$ MeV ($mu= 2.2$ MeV,$\; m_d= 4.7 $MeV) where the latter set is used to estimate the uncertainty of our results due to the uncertainty in the precise values of 
the respective short-distance quark masses. For a systematic discussion of the effects of ISR and IFI as predicted by {\KK}MC-hh on the angular observables in Z production and decay to lepton pairs at the LHC see Ref.~\cite{syost-rdcr19}. Here, we focus on the use of the Z production observables to quantify the systematics in the world-leading ATLAS measurement~\cite{atlasmw-17} of $M_W$:
{\small
$$M_W     =   80370 \pm 7 (\text{stat.}) \pm 11 (\text{exp. syst.}) \pm 14 (\text{mod. syst.}) \text{MeV}\\
=   80370 \pm 19 \text{MeV},$$}
where the first uncertainty is statistical, the second is the experimental systematic uncertainty, and the third is the physics-modeling systematic uncertainty. Z/$\gamma^*$ production
with decay to lepton pairs are used in determining the latter uncertainty. Specifically, the distributions for the lepton $p_T$ spectra in single Z/$\gamma^*$ in Ref.~\cite{atlasmw-17}, as reproduced here in Fig.~\ref{fig2}, are used to analyze expectations for the corresponding lepton spectra in the single W production. 
In both spectra featured in Fig.~\ref{fig2}, we see that, at the lower end, the theory used by ATLAS is about 2\% below the data and, at the high end, it is a similar amount above the data. We address these spectra in Ref.~\cite{kkmchh1} as shown here in Fig.~\ref{fig3}.
 In Fig.~\ref{fig3}, the best prediction of the exact ${\cal O}(\alpha^2 L)$ CEEX calculation (labeled CEEX2) is compared with less precise predictions denoted as follows:
${\cal O}(\alpha^2 L)$ CEEX with ISR+FSR+IFI -- labeled as ``CEEX2''; ${\cal O}(\alpha^2 L)$ CEEX without IFI (initial state final state interference) -- labeled `` CEEX2 (no IFI)'';${\cal O}(\alpha)$ EEX -- labeled ``EEX1''; ${\cal O}(\alpha)$  EEX without ISR (initial state radiation) -- labeled ``EEX1 (no ISR)''.
Here, EEX denotes YFS exponentiation at the level of squared amplitudes as originally formulated in Ref.~\cite{yfs:1961} in contrast to CEEX, which  represents the coherent realization of YFS exponentiation at the level of the amplitudes, as explained in Ref.~\cite{ceex2:1999}. We see that the EEX1 result with no ISR is about 2\% below the best calculation on the low end of the spectrum and is a similar amount above the best calculation on the high end, just as we see in the ATLAS data, where the theory prediction also has no QED ISR other than that in its Pythia~\cite{Sjostrand:2007gs} shower, which is essentially collinear by construction. We conclude that the corrections in {\KK}MC-hh should be implemented into the ATLAS analysis and other similar analyses.\par
\section{Applications in Precision FCC Physics}
The FCC project involves a 100 km tunnel in which one first intalls an $e^+ e^-$ collider that operates for precision Z, WW, Higgs and top physics,to be followed by the installation of a 100 TeV proton-proton collider. The first phase is FCC-ee, the second phase is FCC-hh. As three of us (BFLW, ZAW and SAY) have argued in  Ref.~\cite{radcor17}, IR-improvement of even the FCC-hh discovery spectra is
needed. In the FCC-ee, more than 5 Tera Z's will be observed\footnote{For this scenario, recently, thanks to the modular program design improvements for the electroweak sector calculations became available in {\KK}MC~\cite{ceex2:1999}. The comparison of numerical results is assured. This is of importance for future long-term projects as discussed in \cite{Banerjee:2019mzj}. In particular the up-to-date parametrization of $\alpha_{QED,eff}$\cite{Blondel:2019vdq} (section
B.1) is available as well as the electroweak library {\tt DIZET} version 6.42  \cite{Arbuzov:2005ma}  and the recent version 6.45. The same is true and is of importance for precision measurements of electroweak effects with $\tau$ lepton polarization \cite{Banerjee:2019mle} where benefits from future Belle experimental results on $\tau$ lepton decays are of a sizable potential. In this context technical tests embedded in the older version of {\tt TAUOLA} for the $\tau$ lepton decay \cite{Jadach:1993hs} may need to be revisited to assure technical correctness at better than the precision standard of today.}, so that a key issue is the theoretical precision of the
luminosity as discussed by five of us (SJ, WP, MS, BFLW, SAY) in Ref.~\cite{Jadach:2018}, a discussion to which we now turn.\par
To set the perspective, we show in Tab.~\ref{tab1} the state of the art at the time of LEP, where the precision tag on the theoretical prediction of the luminosity was 0.061\%(0.054\%),
\begin{table}[!ht]
\centering
\scalebox{.85}{
\begin{tabular}{|l|l|l|l|l|l|}
\hline 
    & \multicolumn{2}{|c|}{LEP1} 
              & \multicolumn{2}{|c|}{LEP2}
\\ \hline 
Type of correction/error
    & 1996
         & 1999
              & 1996
                   & 1999
\\  \hline 
(a) Missing photonic ${\cal O}(\alpha^2 )$~\cite{Jadach:1995hy,Jadach:1999pf} 
    & 0.10\%      
        & 0.027\%    
            & 0.20\%  
                & 0.04\%
\\ 
(b) Missing photonic ${\cal O}(\alpha^3 L_e^3)$~\cite{Jadach:1996ir} 
    & 0.015\%     
        & 0.015\%    
            & 0.03\%  
                & 0.03\% 
\\ 
(c) Vacuum polarization~\cite{BW9,BW10} 
    & 0.04\%      
        & 0.04\%    
           & 0.10\%  
                & 0.10\% 
\\ 
(d) Light pairs~\cite{BW11,BW12} 
    & 0.03\%      
        & 0.03\%    
            & 0.05\%  
                & 0.05\% 
\\ 
(e) $Z$ and $s$-channel $\gamma$~\cite{BW13,BW6}
    & 0.015\%      
        & 0.015\%   
            &  0.0\%  
                & 0.0\% 
\\ \hline 
Total  
    & 0.11\%~\cite{BW6}
        & 0.061\%~\cite{bhlumi-precision:1998}
            & 0.25\%~\cite{BW6}
                & 0.12\%~\cite{bhlumi-precision:1998}
\\ \hline 
\end{tabular}}
\caption{\sf
Summary of the total (physical+technical) theoretical uncertainty
for a typical calorimetric detector.
For LEP1, the above estimate is valid for a generic angular range
within   $1^{\circ}$--$3^{\circ}$ ($18$--$52$ mrads), and
for  LEP2 energies up to $176$~GeV and an
angular range within $3^{\circ}$--$6^{\circ}$.
Total uncertainty is taken in quadrature.
Technical precision included in (a).
}
\label{tab1}
\end{table}
according to whether one does not(does) treat the effects of soft pairs~\cite{ON1,ON2}. See Ref.~\cite{bhlumi-precision:1998} for a discussion of the various error estimates as they relate to the cited published works in the table.  The current state of the art, 0.038\%, is shown in Table~\ref{tab2}, where we note that recently this precision tag has been effected
\begin{table}[ht!]
\centering
\scalebox{.85}{
\begin{tabular}{|l|l|l|l|}
\hline
Type of correction~/~Error
    &  1999
        & Update 2018
\\ \hline 
(a) Photonic ${\cal O}(L_e\alpha^2 )$
    & 0.027\% ~\cite{Jadach:1999pf}
        & 0.027\%
\\ 
(b) Photonic ${\cal O}(L_e^3\alpha^3)$
    & 0.015\%~\cite{Jadach:1996ir}
        & 0.015\%
\\ 
(c) Vacuum polariz.
    &0.040\%~\cite{BW9,BW10} 
        & 0.013\%~\cite{JegerlehnerCERN:2016}
\\ 
(d) Light pairs
    & 0.030\%~\cite{BW12}
        & 0.010\%~\cite{ON1,ON2}
\\ 
(e) $Z$ and $s$-channel $\gamma$ exchange
    &0.015\%~\cite{Jadach:1995hv,Arbuzov:1996eq}
        & 0.015\%
\\ 
(f) Up-down interference
    &0.0014\%~\cite{Jadach:1990zf}
        & 0.0014\%
\\ 
(f) Technical Precision& -- & (0.027)\%
\\ \hline  
Total
    & 0.061\%~\cite{bhlumi-precision:1998}
        & 0.038\%
\\ \hline  
\end{tabular}}
\caption{\sf
Summary of the total (physical+technical) theoretical uncertainty for a typical
calorimetric LEP luminosity detector within the generic angular range
of $18$--$52$\,mrad.
Total error is summed in quadrature.
}
\label{tab2}
\end{table}
in Ref.~\cite{pjnt-stj-2019} in reducing the error on the number, $N_\nu$, of light neutrinos so that it is now $N_\nu = 2.9975\pm 0.00742.$ 
The FCC-ee requirement~\cite{fccwksp2019} on the theoretical precision of the luminosity is 0.01\%. The steps involved  in realizing this requirement are discussed in Ref.~\cite{Jadach:2018}.
The conclusion of Ref.~\cite{Jadach:2018} is that, with sufficient resources to make the required improvements to the MC \bhlumi~\cite{bhlumi4:1996}, the 0.01\% precision tag can be realized, as we show in Table~\ref{tab3}. Again, we cite the published works associated to the error estimates in the table as discussed in Ref.~\cite{Jadach:2018}.
\begin{table}[ht!]
\centering
\scalebox{.85}{
\begin{tabular}{|l|l|l|l|}
\hline
Type of correction~/~Error
        & Update 2018
                &  FCC-ee forecast
\\ \hline 
(a) Photonic $[{\cal O}(L_e\alpha^2 )]\; {\cal O}(L_e^2\alpha^3)$
        & 0.027\%
                &  $ 0.1 \times 10^{-4} $
\\ 
(b) Photonic $[{\cal O}(L_e^3\alpha^3)]\; {\cal O}(L_e^4\alpha^4)$
        & 0.015\%
                & $ 0.6 \times 10^{-5} $
\\
(c) Vacuum polariz.
        & 0.014\%~\cite{JegerlehnerCERN:2016}
                & $ 0.6 \times 10^{-4} $
\\
(d) Light pairs
        & 0.010\%~\cite{ON1,ON2}
                & $ 0.5 \times 10^{-4} $
\\
(e) $Z$ and $s$-channel $\gamma$ exchange
        & 0.090\%~\cite{BW13}
                & $ 0.1 \times 10^{-4} $
\\ 
(f) Up-down interference
    &0.009\%~\cite{Jadach:1990zf}
        & $ 0.1 \times 10^{-4} $
\\
(f) Technical Precision & (0.027)\% 
                & $ 0.1 \times 10^{-4} $
\\ \hline 
Total
        & 0.097\%
                & $ 1.0 \times 10^{-4} $
\\ \hline 
\end{tabular}}
\caption{\sf
Anticipated total (physical+technical) theoretical uncertainty 
for a FCC-ee luminosity calorimetric detector with
the angular range being $64$--$86\,$mrad (narrow), near the $Z$ peak.
Description of photonic corrections in square brackets is related to 
the 2nd column.
The total error is summed in quadrature.
}
\label{tab3}
\end{table}
The current planning for the FCC~\cite{fccwksp2019,fccwksp2018}
includes a provision for such resources as those called for in Ref.~\cite{Jadach:2018}.\par
\section{Applications in Quantum Gravity}
We turn now to the role of IR-improvement in quantum gravity. The question of whither or not quantum gravity is calculable in relativistic quantum field theory is still open in the literature, as 
one of us (BFLW) has discussed in Refs.~\cite{rqg-ichep18,ijmpa2018}. Here, one of us argues 
that it is if he extends the YFS~\cite{yfs:1961,yfs1:1988,ceex2:1999} version\footnote{YFS-type soft resummation and its extension to quantum gravity was also worked-out by Weinberg in Ref.~\cite{sw-sftgrav}.} of the exact resummation example to resum the Feynman series for the Einstein-Hilbert Lagrangian for quantum gravity.  The resultant  resummed theory, resummed quantum gravity (RQG), in analogy with what we see in the example in Eq.(\ref{eq-elmtry}), is very much better behaved in the UV compared to what one would estimate from that Feynman series.\par
One of us (BFLW) has shown~\cite{bw1rqg,bw2rqg,bw2arqg,bw2irqg} that the RQG realization of quantum gravity leads to Weinberg's~\cite{wein1} UV-fixed-point behavior for the dimensionless
gravitational and cosmological constants and that the resummed theory is actually UV finite. He has shown further that the RQG theory, taken together with the Planck scale inflationary~\cite{guth,linde} cosmology formulation in Refs.~\cite{reuter1,reuter2}\footnote{The authors in Ref.~\cite{sola1} also proposed the attendant 
choice of the scale $k\sim 1/t$ used in Refs.~\cite{reuter1,reuter2}.} from the 
asymptotic safety approach to quantum gravity in 
Refs.~\cite{reutera,laut,reuterb,reuter3,litim,litim1,litim2,perc,perc1,perc2,perc3,perc4}, allows us to predict the cosmological constant $\Lambda$. Specifically, 
he obtains, employing the arguments in Refs.~\cite{branch-zap} ($t_{eq}$ is the time of matter-radiation equality),  the result~\cite{drkuniv}{\small
\begin{equation}
\begin{split}
\rho_\Lambda(t_0)&\cong \frac{-M_{Pl}^4(1+c_{2,eff}k_{tr}^2/(360\pi M_{Pl}^2))^2}{64}\sum_j\frac{(-1)^Fn_j}{\rho_j^2}
          \times \frac{t_{tr}^2}{t_{eq}^2} \times (\frac{t_{eq}^{2/3}}{t_0^{2/3}})^3\cr
   & \cong \frac{-M_{Pl}^2(1.0362)^2(-9.194\times 10^{-3})}{64}\frac{(25)^2}{t_0^2}
   \cong (2.4\times 10^{-3}eV)^4.\cr
\end{split}
\label{eq-rho-expt}
\end{equation}}
$t_0\cong 13.7\times 10^9$ yrs  is the age of the universe, $t_{tr}\sim 25 t_{Pl}$ is the transition time between the Planck regime and the classical Friedmann-Robertson-Walker(FRW) regime in 
the Planck scale cosmology description of inflation in Ref.~\cite{reuter2}, $c_{2,eff}\cong  2.56\times 10^4$ is defined in Refs.~\cite{bw1rqg,bw2rqg,bw2arqg}, $t_{eq}$ is the time of matter-radiation equality, and $M_{Pl}$ is the Planck mass.\par
Due to the prediction's closeness to the observed value~\cite{cosm1a,cosm1b,pdg2008},$\rho_\Lambda(t_0)|_{\text{expt}}\cong ((2.37\pm 0.05)\times 10^{-3}eV)^4$, one of us (BFLW) discusses in Ref.~\cite{eh-consist} its reliability and argues that its uncertainty is at the level of a factor of ${\cal O}(10)$. There follow constraints on susy GUT's as well as 
discussed in Refs.~\cite{drkuniv}. Various consistency checks on our result for $\rho_\Lambda(t_0)$ are also considered in Ref.~\cite{drkuniv}.\par
\section{Summary}
What we can see is that the resummation of the IR regime of quantum field theory, coupled with exact results to a given order, when done at the level of the amplitude in an exact
re-arrangement of the original Feynman series, has wide applicability for precision phenomenology. We have illustrated the range of this applicability from the current precision LHC physics to the futuristic FCC precision physics program and reliable estimates for quantum gravity by one of us (BFLW). The future of particle physics is intimately interwoven with continued progress on IR-improved quantum field theoretic predictions, when taken together with control of the attendant collinear and UV limits. \par 

\vskip 2 mm
\centerline{\bf Acknowledgments}
\vskip 2 mm

This work was supported in part
by the Programme of the French–Polish Cooperation between IN2P3 and COPIN 
within the Collaborations Nos. 10-138 and 11-142 and by a grant from the Citadel Foundation. The authors also thank Prof. G. Giudice for the support and kind hospitality of the CERN TH Department. 
 
\bibliography{Tauola_interface_design}{}

\providecommand{\href}[2]{#2}\begingroup\begin{thebibliography}{100}

\bibitem{djg-smat50}
D.~J. Gross, {in {\it Proc. SM@50}, Cambridge University Press, 2019, in
  press}.

\bibitem{frits-ichep88}
F.~A. Berends, {private communication, 1988}.

\bibitem{HERWIG}
G.~Corcella, I.~G. Knowles, G.~Marchesini, S.~Moretti, K.~Odagiri,
  P.~Richardson, M.~H. Seymour, and B.~R. Webber, {\em JHEP} {\bf 0101} (2001)
  010, \href{http://www.arXiv.org/abs/hep-ph/0011363}{{\tt hep-ph/0011363}}.

\bibitem{Bahr:2008pv}
M.~Bahr {\em et al.}, {\em Eur. Phys. J.} {\bf C58} (2008) 639--707,
\href{http://www.arXiv.org/abs/0803.0883}{{\tt 0803.0883}}.

\bibitem{Pythia}
{T. Sjostrand} {\em et al.}, {\em Comput. Phys. Commun.} {\bf 135} (2001) 238.

\bibitem{sherpa}
T.~Gleisberg {\em et al.}, {\em J. High Energy Phys.} {\bf 02} (2009) 007,
  \href{http://www.arXiv.org/abs/hep-ph/0811.4622}{{\tt hep-ph/0811.4622}}.

\bibitem{hwiri}
{S. Joseph} {\em et al.}, {\em Phys. Rev. D} {\bf 81} (2010) 076008,
  \href{http://www.arXiv.org/abs/hep-ph/1001.1434}{{\tt hep-ph/1001.1434}}.

\bibitem{qedpdf}
A.~Manohar {\em et al.}, {\em Phys. Rev. Lett.} {\bf 117} (2016) 242002.

\bibitem{kkmchh}
S.~Jadach, B.~F.~L. Ward, Z.~W\c{a}s, and S.~Yost, {\em Phys. Rev. D} {\bf 94}
  (2016) 074006, \href{http://www.arXiv.org/abs/hep-ph/1608.01260}{{\tt
  hep-ph/1608.01260}}.

\bibitem{reutera}
M.~Reuter, {\em Phys. Rev. D} {\bf 57} (1998) 971.

\bibitem{laut}
O.~Lauscher and M.~Reuter, {\em Phys. Rev. D} {\bf 66} (2002) 025026.

\bibitem{reuterb}
E.~Manrique, M.~Reuter, and Saueressig, {\em Ann. Phys.} {\bf 326} (2011) 44.

\bibitem{reuter3}
A.~Bonanno and M.~Reuter, {\em Phys. Rev. D} {\bf 62} (2000) 043008.

\bibitem{litim}
{ D. F. Litim}, {\em Phys. Rev. Lett.} {\bf 92} (2004) 201301.

\bibitem{litim1}
{ D. F. Litim}, {\em Phys. Rev. D} {\bf 64} (2001) 105007.

\bibitem{litim2}
{P. Fischer and D. F. Litim}, {\em Phys. Lett. B} {\bf 638} (2006) 497.

\bibitem{perc}
{ D. Don and R. Percacci, Class. Quant. Grav. {\bf 15} (1998) 3449}, {\em
  Class. Quant. Grav.} {\bf 15} (1998) 3449.

\bibitem{perc1}
{R. Percacci and D. Perini}, {\em Phys. Rev. D} {\bf 67} (2003) 081503.

\bibitem{perc2}
{ R. Percacci}, {\em Phys. Rev. D} {\bf 73} (2006) 041501.

\bibitem{perc3}
{R. Percacci and D. Perini}, {\em Phys. Rev. D} {\bf 68} (2003) 044018.

\bibitem{perc4}
{ A. Codello and R. Percacci and C. Rahmede}, {\em Int.J. Mod. Phys. A} {\bf
  23} (2008) 143.

\bibitem{bw1rqg}
{ B.F.L. Ward}, {\em Open Nucl.Part.Phys.Jour.} {\bf 2} (2009) 1.

\bibitem{bw2rqg}
{ B.F.L. Ward}, {\em Mod. Phys. Lett. A} {\bf 17} (2002) 237.

\bibitem{bw2arqg}
{B.F.L. Ward}, {\em Mod. Phys. Lett.} {\bf 19} (2004) 143.

\bibitem{bw2irqg}
{B.F.L. Ward}, {\em Mod. Phys. Lett. A} {\bf 23} (2008) 3299.

\bibitem{SM3}
S.~L. Glashow, {\em Nucl. Phys.} {\bf 22} (1961) 579--588.

\bibitem{SM4}
A.~Salam, {\em Elementary Particle Theory}.
\newblock N. Svartholm (Almqvist and Wiksell), Stockholm, 1968.

\bibitem{SM1}
S.~Weinberg, {\em Phys. Rev. Lett.} {\bf 19} (1967) 1264--1266.

\bibitem{BW1a}
G.~'t~Hooft and M.~Veltman, {\em Nucl. Phys. B} {\bf 44} (1972) 189.

\bibitem{BW1b}
D.~'Gross and F.~Wilczek, {\em Phys. Rev. Lett.} {\bf 30} (1973) 1343.

\bibitem{BW1c}
H.~D. 'Politzer, {\em Phys. Rev. Lett.} {\bf 30} (1973) 1346.

\bibitem{case-sm50}
B.~W. Lynn {\em et al.}, {{\it Proc. SM@50}, Cambridge University Press, 2019,
  in press}.

\bibitem{slacsi2018}
T.~Rizzo {\em et al.}, {{\it Proc. 46th SLAC Summer Institute}, C18-07-30,
  2018}.

\bibitem{fcc-fabiola-2019}
F.~Gianotti, {in {\it New Year presentation}, CERN, Geneva, Switzerland, Jan.
  15, 2019}.

\bibitem{mcnlo-hwiri}
{ S.K. Majhi} {\em et al.}, {\em Phys. Lett. B} {\bf 719} (2013) 367,
  \href{http://www.arXiv.org/abs/hep-ph/1208.4750}{{\tt hep-ph/1208.4750}}.

\bibitem{mcnlo-hwiri1}
{ A. Mukhopadhyay and B.F.L. Ward}, {\em Mod. Phys. Lett. A} {\bf 31} (2016)
  1650063, \href{http://www.arXiv.org/abs/hep-ph/1412.8717}{{\tt
  hep-ph/1412.8717}}.

\bibitem{mcnlo}
{S. Frixione and B.Webber}, {\em J. High Energy Phys.} {\bf 0206} (2002) 029.

\bibitem{mcnlo1}
{ S. Frixione} {\em et al.}, {\em J. High Energy Phys.} {\bf 1101} (2011) 053,
  \href{http://www.arXiv.org/abs/hep-ph/1010.0568}{{\tt hep-ph/1010.0568}}.

\bibitem{mg5_amc-hwri}
{ B. Shakerin and B.F.L. Ward}, {\em Phys. Rev. D} {\bf 100} (2019) 034026,
  \href{http://www.arXiv.org/abs/hep-ph/1809.01492}{{\tt hep-ph/1809.01492}}.

\bibitem{dglap1}
G.~Altarelli and G.~Parisi, {\em Nucl. Phys. B} {\bf 126} (1977) 298.

\bibitem{dglap2}
Y.~L. Dokshitzer, {\em Sov. Phys. JETP} {\bf 46} (1977) 641.

\bibitem{dglap3}
L.~Lipatov, {\em Yad. Fiz.} {\bf 20} (1974) 181.

\bibitem{dglap4}
V.~Gribov and L.~Lipatov, {\em Sov. J. Nucl. Phys.} {\bf 15} (1972) 675, 938.

\bibitem{dglap5}
J.~Collins and J.~Qiu, {\em Phys. Rev. D} {\bf 39} (1989) 1398.

\bibitem{dglap6}
C.~G. Callan~Jr., {\em Phys. Rev. D} {\bf 2} (1970) 1541.

\bibitem{dglap7}
K.~Symanzik, {\em Commun. Math. Phys.} {\bf 18} (1970) 227.

\bibitem{dglap8}
K.~Symanzik, {\em Springer Tracts. Mod. Phys.} {\bf 57} (1971) 222.

\bibitem{dglap9}
S.~Weinberg, {\em Phys. Rev. D} {\bf 8} (1973) 3497.

\bibitem{mg5amcnlo}
J.~Alwall {\em et al.}, {\em J. High Energy Phys.} {\bf 07} (2014) 079,
  \href{http://www.arXiv.org/abs/hep-ph/1405.0301}{{\tt hep-ph/1405.0301}}.

\bibitem{kkmchh1}
S.~Jadach, B.~F.~L. Ward, Z.~W\c{a}s, and S.~Yost, {\em Phys. Rev. D} {\bf 99}
  (2019) 076016, \href{http://www.arXiv.org/abs/hep-ph/1707.06502}{{\tt
  hep-ph/1707.06502}}.

\bibitem{radcor17}
{B.F.L. Ward} {\em et al.}, {\em PoS} {\bf RADCOR2017} (2018)
083.

\bibitem{lhe-formt}
{E. Boos} {\em et al.}, \href{http://www.arXiv.org/abs/hep-ph/0109068}{{\tt
  hep-ph/0109068}}.

\bibitem{zfitter1}
D.~Bardin {\em et al.}, EW Library.

\bibitem{zfitter6:1999}
D.~Bardin {\em et al.},
e-print: hep-ph/9908433.

\bibitem{vicini-wack:2016}
S.~Alioli {\em et al.}, {\em CERN-TH-2016-137; CERN-LPCC-2016-002} (2016)
  \href{http://www.arXiv.org/abs/hep-ph/1606.02330}{{\tt hep-ph/1606.02330}}.

\bibitem{syost-rdcr19}
S.~A. Yost {\em et al.}, {in {\it Proc. RADCOR2019}, PoS, 2019, in press}.

\bibitem{atlasmw-17}
{ATLAS} Collaboration, M.~Aaboud {\em et al.}, {\em Eur. Phys. J. C} {\bf 78}
  (2018) 110, \href{http://www.arXiv.org/abs/hep-ex/1701.07240}{{\tt
  hep-ex/1701.07240}}.

\bibitem{yfs:1961}
D.~R. Yennie, S.~Frautschi, and H.~Suura, {\em Ann. Phys. (NY)} {\bf 13} (1961)
  379 -- 452.

\bibitem{ceex2:1999}
S.~Jadach, B.~F.~L. Ward, and Z.~W\c{a}s, {\em Phys. Rev. D} {\bf 63} (2001)
  113009, preprint UTHEP-99-09-01.

\bibitem{Sjostrand:2007gs}
T.~Sjostrand, S.~Mrenna, and P.~Skands, {\em Comput. Phys. Commun.} {\bf 178}
  (2008) 852--867,
\href{http://www.arXiv.org/abs/0710.3820}{{\tt 0710.3820}}.

\bibitem{Banerjee:2019mzj}
S.~Banerjee, M.~Chrzaszcz, Z.~Was, and J.~Zaremba, ``{Heritage projects,
  archivization and re-usability concerns}'', in {\em {Theory report on the
  11th FCC-ee workshop}}, pp.~137--140,
2019.

\bibitem{Blondel:2019vdq}
{\em {Theory report on the 11th FCC-ee workshop}},
2019.
\newblock

\bibitem{Arbuzov:2005ma}
A.~B. Arbuzov, M.~Awramik, M.~Czakon, A.~Freitas, M.~W. Grunewald, K.~Monig,
  S.~Riemann, and T.~Riemann, {\em Comput. Phys. Commun.} {\bf 174} (2006)
  728--758,
\href{http://www.arXiv.org/abs/hep-ph/0507146}{{\tt hep-ph/0507146}}.

\bibitem{Banerjee:2019mle}
S.~Banerjee and Z.~Was, ``{FCC Tau Polarization}'', in {\em {Theory report on
  the 11th FCC-ee workshop}}, pp.~215--216,
2019.

\bibitem{Jadach:1993hs}
S.~Jadach, Z.~W\c{a}s, R.~Decker, and J.~H. K\"{uhn}, {\em Comput. Phys.
  Commun.} {\bf 76} (1993)
361.

\bibitem{Jadach:2018}
S.~Jadach, W.~Placzek, M.~Skrzypek, B.~F.~L. Ward, and S.~A. Yost, {\em Phys.
  Lett. B} {\bf 790} (2018) 314.

\bibitem{Jadach:1995hy}
S.~Jadach, M.~M.~Melles, B.~F.~L. Ward, and S.~A. Yost, {\em Phys. Lett.} {\bf
  B377} (1996) 168--176.

\bibitem{Jadach:1999pf}
S.~Jadach, M.~M.~Melles, B.~F.~L. Ward, and S.~A. Yost, {\em Acta Phys. Polon.}
  {\bf B30} (1999)
1745 -- 1750.

\bibitem{Jadach:1996ir}
S.~Jadach and B.~F.~L. Ward, {\em Phys. Lett.} {\bf B389} (1996)
129--136.

\bibitem{BW9}
H.~Burkhardt and B.~Pietrzyk, {\em Phys. Lett. B} {\bf 356} (1995)
398 -- 403.

\bibitem{BW10}
S.~Eidelman and F.~F.~Jegerlehner, {\em Z. Phys. C} {\bf 67} (1995)
585 -- 602.

\bibitem{BW11}
S.~Jadach, M.~Skrzypek, and B.~Ward, {\em Phys. Rev. D} {\bf 47} (1993)
3733 -- 3741.

\bibitem{BW12}
S.~Jadach, M.~Skrzypek, and B.~Ward, {\em Phys. Rev. D} {\bf 55} (1997)
1206 -- 1215.

\bibitem{BW13}
S.~Jadach, W.~Placzek, and B.~Ward, {\em Phys. Lett. B} {\bf 353} (1995)
349 -- 361.

\bibitem{BW6}
A.~Arbuzov {\em et al.}, {\em Phys. Lett. B} {\bf 383} (1996) 238 -- 242,
\href{http://www.arXiv.org/abs/hep-ph/9605239}{{\tt hep-ph/9605239}}.

\bibitem{bhlumi-precision:1998}
S.~Jadach, M.~M.~Melles, B.~F.~L. Ward, and S.~A. Yost, {\em Phys. Lett.} {\bf
  B450} (1999) 262--266.

\bibitem{ON1}
G.~Montagna {\em et al.}, {\em Nucl. Phys. B} {\bf 547} (1999) 39 -- 59,
\href{http://www.arXiv.org/abs/hep-ph/9811436}{{\tt hep-ph/9811436}}.

\bibitem{ON2}
G.~Montagna {\em et al.}, {\em Phys. Lett.} {\bf B459} (1999) 649 -- 652,
\href{http://www.arXiv.org/abs/hep-ph/9905235}{{\tt hep-ph/9905235}}.

\bibitem{JegerlehnerCERN:2016}
F.~Jegerlehner, talk in FCC-ee Mini-Workshop, {\em Physics Behind Precision},
  {\small
  https://indico.cern.ch/event/469561/contributions/1977974/attachments/1221704/1786449/SMalphaFCCee16.pdf}.

\bibitem{Jadach:1995hv}
S.~Jadach, W.~Placzek, and B.~F.~L. Ward, {\em Phys. Lett.} {\bf B353} (1995)
349--361.

\bibitem{Arbuzov:1996eq}
A.~Arbuzov {\em et al.}, {\em Phys. Lett.} {\bf B383} (1996) 238--242,
\href{http://www.arXiv.org/abs/hep-ph/9605239}{{\tt hep-ph/9605239}}.

\bibitem{Jadach:1990zf}
S.~Jadach, E.~E.~Richter-W\c{a}s, B.~F.~L. Ward, and Z.~W\c{a}s, {\em Phys.
  Lett.} {\bf B253} (1991)
469 -- 477.

\bibitem{pjnt-stj-2019}
P.~Janot and S.~Jadach, \href{http://www.arXiv.org/abs/hep-ph/1912.02067}{{\tt
  hep-ph/1912.02067}}.

\bibitem{fccwksp2019}
A.~Blondel {\em et al.}, \href{http://www.arXiv.org/abs/hep-ph/1905.05078}{{\tt
  hep-ph/1905.05078}}.

\bibitem{bhlumi4:1996}
S.~Jadach, W.~Placzek, E.~Richter-W\c{a}s, B.~F.~L. Ward, and Z.~W\c{a}s, {\em
  Comput. Phys. Commun.} {\bf 102} (1997)
229.

\bibitem{fccwksp2018}
A.~Blondel {\em et al.}, \href{http://www.arXiv.org/abs/hep-ph/1809.01830}{{\tt
  hep-ph/1809.01830}}.

\bibitem{rqg-ichep18}
{B.F.L. Ward}, {\em PoS} {\bf ICHEP2018} (2018)
620.

\bibitem{ijmpa2018}
{B.F.L. Ward}, {\em Int. J. Mod. Phys. A} {\bf 33} (2018) 1830028.

\bibitem{yfs1:1988}
S.~Jadach and B.~F.~L. Ward, {\em Phys. Rev.} {\bf D38} (1988) 2897.

\bibitem{sw-sftgrav}
S.~Weinberg, {\em Phys. Rev.} {\bf 140} (1965) B516.

\bibitem{wein1}
S.~Weinberg, {in {\it General Relativity, an Einstein Centenary Survey}, eds.
  S. W. Hawking and W. Israel, (Cambridge Univ. Press, Cambridge, 1979)}.

\bibitem{guth}
A.~H. Guth, {\em {Phys. Rev.D}} {\bf 23} (1981) 347.

\bibitem{linde}
A.~Linde, {\em {Lecture Notes in Phys.}} {\bf 738} (2008) 1.

\bibitem{reuter1}
{A. Bonanno and M. Reuter}, {\em Phys. Rev. D} {\bf 65} (2002) 043508.

\bibitem{reuter2}
{A. Bonanno and M. Reuter}, {\em Jour. Phys. Conf. Ser.} {\bf 140} (2008)
  012008.

\bibitem{sola1}
I.~L. Shapiro and J.~Sola, {\em Phys. Lett. B} {\bf 475} (2000) 236.

\bibitem{branch-zap}
{V. Branchina and D. Zappala}, {\em G. R. Gravit.} {\bf 42} (2010) 141.

\bibitem{drkuniv}
{B.F.L. Ward}, {\em Mod. Phys. Lett. A} {\bf 30} (2015) 1550206.

\bibitem{cosm1a}
A.~G. Riess {\em et al.}, {\em Astron. Jour.} {\bf 116} (1998) 1009.

\bibitem{cosm1b}
S.~Perlmutter {\em et al.}, {\em Astrophys. Jour.} {\bf 517} (1999) 565.

\bibitem{pdg2008}
C.~Amsler {\em et al.}, {\em Phys. Lett. B} {\bf 667} (2008) 1.

\bibitem{eh-consist}
{B.F.L. Ward}, {\em Phys. Rev. D} {\bf 57} (1998) 971.

\end{thebibliography}\endgroup
\bibliographystyle{utphys_spires}


\end{document}